\documentclass[journal, 19pt]{IEEEtranTIE}
\usepackage{graphicx}
\usepackage{cite}
\usepackage{picinpar}
\usepackage{amsmath}
\usepackage{url}
\usepackage{flushend}
\usepackage[latin1]{inputenc}
\usepackage{colortbl}
\usepackage{soul}
\usepackage{multirow}
\usepackage{pifont}
\usepackage{color}
\usepackage{alltt}
\usepackage[hidelinks]{hyperref}
\usepackage{enumerate}
\usepackage{siunitx}
\usepackage{breakurl}
\usepackage{algorithm}
\usepackage{algpseudocode}
\usepackage{epstopdf}
\usepackage{pbox}
\usepackage{caption}
\usepackage{subcaption}
\begin{document}
\title{Design and Implementation of DC-to-5~MHz Wide-Bandwidth High-Power High-Fidelity Converter}



\author{
	\vskip 1em
	
    Jinshui Zhang, \emph{Student Member},
	Boshuo Wang, \emph{Member},
	Xiaoyang Tian, \emph{Member},
    Angel Peterchev, \emph{Member},
	and Stefan Goetz, \emph{Member}
}

\maketitle
	
\begin{abstract}
    Advances in power electronics have made it possible to achieve high power levels, e.g., reaching GW in grids, or alternatively high output bandwidths, e.g., beyond MHz in communication. Achieving both simultaneously, however, remains challenging. 
    Various applications, ranging from efficient multichannel wireless power transfer to cutting-edge medical and neuroscience applications, are demanding both high power and wide bandwidth. 
    Conventional inverters can achieve high power and high quality at grid or specific frequency ranges but lose their fidelity when reaching higher output frequencies. 
    Resonant circuits can promise a high output frequency but only a narrow bandwidth. 
    We overcome the hardware challenges by combining gallium-nitride (GaN) transistors with modular cascaded double-H bridge circuits and control that can manage typical timing and balancing issues.
    We developed a lightweight embedded control solution that includes an improved look-up-table digital synthesizer and a novel adaptive-bias-elimination nearest-level modulation. 
    This solution effectively solves the conflict between a high power level and high output bandwidth and can---in contrast to previous approaches---in principle be scaled in both dimensions.
    Our prototype exhibits a frequency range from DC to 5 MHz with $<18\,\%$ total voltage distortion across the entire frequency spectrum, while achieving a power level of $>5~\mathrm{kW}$.
    We conducted tests by sweeping the output frequency and two channel-mixing trials, which included a practical magnetogenetics-oriented stimulation pulse and an entertaining trial to reproduce the famous \textit{Arecibo} message with the current spectrum.
\end{abstract}

\begin{IEEEkeywords}
Cascaded bridge converter, 
 modular multilevel converter, 
	magnetogenetics,
	high-frequency modulation, frequency-division multiplexing,
	wide output bandwidth, 
	nearest level modulation / control
\end{IEEEkeywords}

{}

\section{Introduction}

The demands for high-power wide-bandwidth power converters have grown rapidly and widely, driven by applications such as wireless power transfer \cite{9912436, 9638501, 8630668,9947537,9169849}, power-hardware-in-the-loop systems \cite{9766331}, efficient wide-bandwidth radio-frequency amplifiers \cite{7855275,5272099,5316317}, scientific instrumentation \cite{9328613,doi:10.1063/5.0046706,7794102}, and recent biomedical applications \cite{hayashi2010high, monsalve2015remotely, oliveira2013magnetic, huang2010remote, chen2015wireless, munshi2017magnetothermal, Wang_2022, DengTechnology}. 
Magnetothermogenetics as a cutting-edge neuronal stimulation technique, for instance, uses magnetic fields to remotely control cell activities and particularly inject neuronal signals through activating electromagnetic receivers in the form of nanoparticle--protein compounds \cite{sebesta2022subsecond}. 
This technology can selectively actuate thermally sensitive ion channel proteins with high-frequency heating of iron oxide nanoparticles, but requires considerable power and freely tunable high-quality output with low distortion for frequency-multiplexed neural stimulation to avoid activating other nanoparticles with side-bands or harmonics.  
Additionally, concurrent control of a wider output bandwidth instead of only limited resonances would allow full frequency-division multiplexing for writing neural signals into the brain or for (optionally encrypted) wireless power transfer \cite{6928497}. 

Wide output bandwidth from DC to megahertz, high quality, and high efficiency at the same time are challenging conventional power electronics.
Typical high-power conversion for electricity grids is best for low frequencies, often as low as the grid frequency, and may offer high power levels but only low bandwidth \cite{7884555,9672538,9940564}. 
Electric powertrains extend the bandwidth only moderately, typically within the single-digit kilohertz range \cite{6231806, 9135026, 6231806, 6342749, 9135026, 6397383}.
Existing solutions to achieving both higher output frequency and high power mainly involve resonant circuits \cite{Wang_2022, sebesta2021sub, 9336630, peterchev2015advances}. 
However, the output bandwidth of these circuits is narrow around the resonant frequency. 
Considerable variation of output frequency requires reconfiguring the hardware \cite{Wang_2022}. 

A promising direction for such research is cascaded bridge (CBC) and modular multilevel converters (MMC). 
These techniques offer a highly scalable and flexible circuit, making them already popular with power delivery and conversion applications, such as high-voltage AC/DC converters \cite{4591920, 5544594,5637505,6038879,6085335} and medium-voltage motor drives \cite{5984128,6397383,5279054,6342749,6231806,8601394}.
Furthermore, cascaded circuits spread voltage, current, and switching across a specified number of modules and semiconductors so that they can increase both the output quality and the power--bandwidth product, promising concurrent high bandwidth, high quality, and high power.
Previous research has already explored their applications for high-quality pulse synthesizing, demonstrating an output bandwidth in the kHz range and power levels in the MVA range \cite{zeng2022modular, li2022modular, wang2022optimized, mps2012circuit}. 
However, existing cascaded circuit techniques alone cannot achieve the required DC-to-MHz bandwidth with sufficient power and high quality. 

On the one hand, transistors and the circuit topology determine the theoretical output bandwidth of converters. 
Aiming for a MHz output profile, gallium-nitride (GaN) transistors stand out due to their fast switching dynamics and low parasitic parameters.
Balancing of the modules, e.g., with respect to voltage, power, and switching, is another crucial consideration. 
The wide bandwidth from DC to MHz rules out most previous options due to their need for module monitoring and sorting, which conflicts with the bandwidth limits of sampling, digital filtering, or sorting algorithms \cite{8029004, 7361408, 8962148, 8722162, 6147159, 8681892}. 
The circuit requires a sensorless and fast-responding voltage-sharing method. 


On the other hand, the embedded control requires improvement in synthesis and modulation of variable high-bandwidth signals. 
Traditional synthesis of sinusoidal signals typically relies on look-up tables (LUT) \cite{8938571, 9545341, 1243689, ana_dds_1999}. 
However, this method leads to bias when facing high-bandwidth output. 
Direct digital synthesizer (DDS) technology addresses this issue by introducing a phase accumulator, which requires a tremendous memory size that exponentially increases with the resolution \cite{ana_dds_1999}. 
With limited block memory in embedded controllers, this method degrades into an approach that inherits errors from each output cycle, causing a low-frequency oscillation, i.e., subharmonic distortion \cite{9888117}.
Besides synthesizing, modulating sinusoidal signals also runs into issues at high frequencies. 
For example, pulse width modulation (PWM) methods, though conventionally popular, struggles with very-high-bandwidth signals due to interactions between the signal and the carrier(s) \cite{holmes2003pulse}.
Instead, nearest-level modulation (NLM) does not rely on any carrier and offers a wide bandwidth close to the switching rate. 
However, due to only few clock cycles covering high-frequency output as well as the limited precision number representation in embedded controllers, NLM may suffer from numerical asymmetry errors caused by rounding operations, leading to DC bias or low-frequency oscillations and potentially resulting in over-current.

To address hardware challenges, we combine the cascaded double-H bridge circuit, which supports a fast sensorless voltage balancing method \cite{7468193, 9109731, 8096011, li2018module, li2020reduced}, with GaN transistors to provide both wide output bandwidth and high power level. 
For the embedded control, we modify the look-up table method to synthesize high-frequency sinusoidal signals in embedded controllers without causing DC bias or oscillation. 
Additionally, we improve the nearest-level modulation to adaptively eliminate the numeric error due to limited precision. 

%
\section{Concept}

\subsection{Circuit}
\begin{figure}
    \centering
    \includegraphics{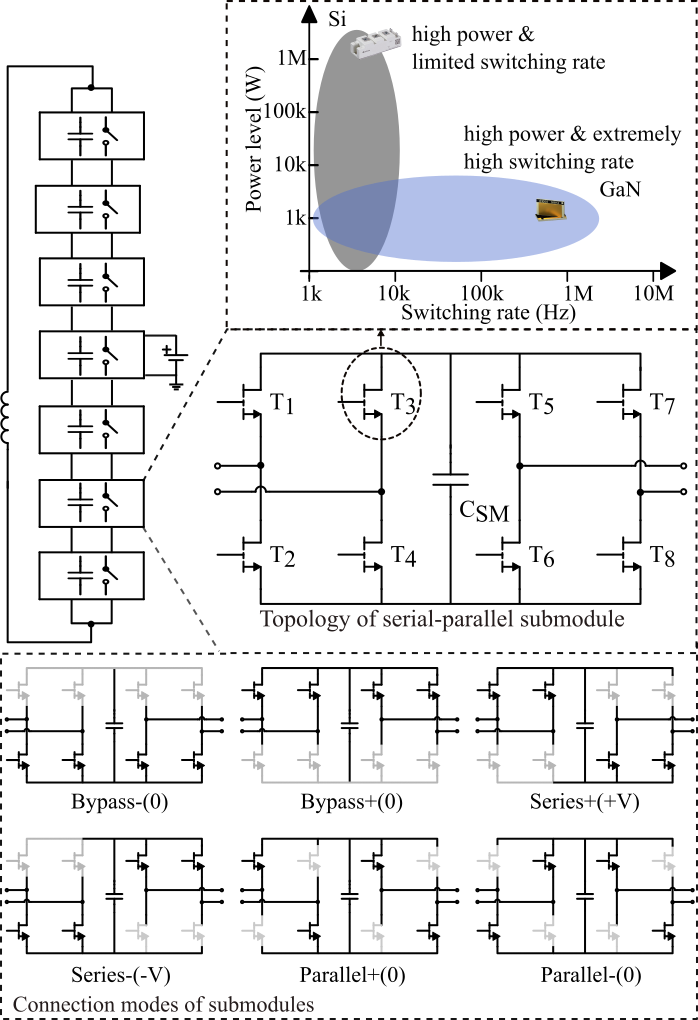}
    \caption{Single-arm demonstration circuit consisting of seven GaN-based parallel-series modules. 
    Each module has two H-bridges and can generate an output level of zero, positive or negative capacitor voltage. 
    The double H-bridge structure enables parallel connectivity and thus ideally balanced voltage among modules. 
    GaN transistors offer high switching rates, enabling theoretically high output frequencies.}
    \label{fig:prototype_topology}
\end{figure}

Figure \ref{fig:prototype_topology} describes the circuit configuration. 
The system consists of seven cascaded double-H bridge submodules.
The submodules can assume six states, including Bypass+, Bypass--, Series+, Series--, Parallel+ and Parallel--, respectively generating an output of \{0, 0, +V, -V, 0, 0\} \cite{9415177, 8601394}. 
Parallel connectivity simplifies the overall topology of the high-voltage DC bus, rendering the necessity for just a single low-voltage DC source, which is then connected to the fourth module as depicted in Figure \ref{fig:prototype_topology} \cite{9415177}. The system architecture incorporates GaN enhancement-mode transistors due to their capability for MHz-level switching rates.

The standout advantage of this topology is its capability for sensorless voltage balancing. The double-H bridge configuration achieves voltage equilibrium across modules through rapid paralleling, resulting in modestly increased switching loss, though \cite{6763109}. This approach ensures efficient energy distribution among modules without necessitating monitoring or sorting procedures, which run into limitations at the required speeds.

On the other hand, it is important to recognize the trade-off between the frequency at which modules clear their imbalance through parallelization, which can drive switching loss, and the energy loss caused by parallelization, if the modules had too much time building up a significant imbalance. A faster paralleling process leads to reduced losses during parallelization, but at the expense of increased switching losses.

\subsection{Embedded Control Design}

\subsubsection{Reference Signal Synthesis}
The initial stage of the control process involves synthesizing the reference signal, which consists of sinusoidal signals with a wide range of frequencies. Embedded controllers typically synthesize sinusoidal signals through a look-up table (LUT). However, when aiming to achieve an output frequency spanning up to 5 MHz, the traditional LUT method falls short of delivering the desired output quality.

\begin{table}
\caption{Quantities Definition}
\label{tab:lut_var_def}
\setlength{\tabcolsep}{3pt}
\centering
\begin{tabular}{p{45pt}|p{140pt}}
\hline
Symbol& Quantities or Operation \\
\hline
	$f_{\mathrm{o}}$& Objective frequency\\
	$f_{\mathrm{clock}}$& FPGA clock rate\\
	$k$ & Index of FPGA clock within output cycle \\
	$K$ & Number of FPGA clocks during each output cycle \\
	$n$ & Index of output cycle \\
	$L$ & Length of look-up table \\
	$A(k)$ & Address of look up table at $k$-th FPGA clock\\
	$A_0(n)$ & Initial address of $n$-th output cycle of look up table \\
	$\Delta A$ & Address increment for each FPGA clock \\
	$ s_{\mathrm{ref, fp}} $ & Reference signal,  fixed-point numbers \\
	$ s_{\mathrm{mod, inte}} $ & Digitized modulation signal, integers \\
\hline
\end{tabular}
\end{table}
Key variables in LUT, including the address increment for each FPGA clock, the number of clock cycles of the controller per output cycle, and the index within an output cycle, respectively follow
\begin{align}
	\begin{split}
    		\Delta A &= L  \frac{f_o}{f_{\mathrm{clock}}} \\
    		K & = \Bigl\lfloor \frac{L}{\Delta A} \Bigr\rceil\\
    		k &\in \{0, 1, \cdots, K-1 \}.
	\end{split}
\end{align}

A sinusoidal reference waveform of predefined length $L$ is stored in the embedded controller. Flip-flops can optionally offer faster reading rates or memory for accommodating larger table sizes. If the output frequency $f_o$ remains constant, we can optimize the length of the look-up table to ensure symmetric table value visits during one output cycle, thereby guaranteeing a zero net accumulation.
Nevertheless, when the output frequency varies, the conventional LUT method falls short of covering the complete frequency range without resulting in DC bias in the output voltage and subsequent excessive over-current.

\begin{figure}
    \centering
    \includegraphics{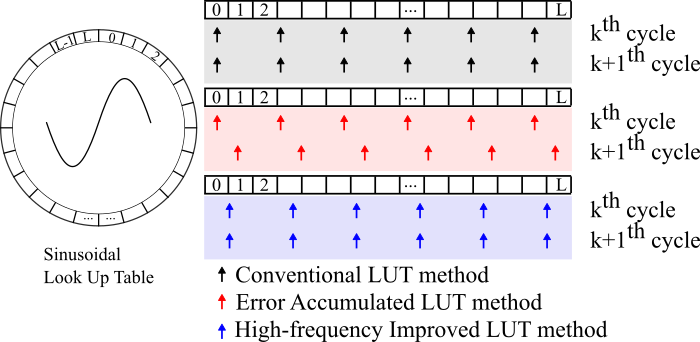}
    \caption{Principles of different LUT methods, respectively, 
    a) The conventional LUT approach, which always visits from zero for every output cycle; 
    b) The error-inherited LUT approach, which visits the table from the bias address obtained from the last output cycle; 
    c) The improved LUT method, which starts from an initial address that is pre-calculated according to the output frequency and avoids DC bias or low-frequency oscillation.}
    \label{fig:lut_methods}
\end{figure}
For every output cycle, the conventional LUT method starts visiting the table at $A_0 = 0$.
However, this method can lead to non-zero summation and thus a DC bias, $\sum A(k) \neq 0$
We can handle DC bias error by starting the address from a non-zero initial address for each output cycle per
\begin{equation}
	A_0(n) = (A_{K}(n-1) + \Delta A) - L,
	\label{equ:inherite_initial_lut_address}
\end{equation}
where $A_{K}(n-1)$ is the last address of the $(n-1)$-th output cycle, $A_{0}(n)$ the first address of the $n$-th output cycle. 
By inheriting errors of the previous cycle, the lookup table will be repeated circularly, and the accumulated net value alternates with a zero average. 
This method can effectively reduce the DC bias and prevent the circuit from damage caused by infinite current in the inductor load.
However, an issue with this LUT method is that it leads to a periodically alternating net value.
Instead of generating a DC bias in the output, there is a low-frequency (LF) oscillation caused by the periodically changing net value for each individual cycle.
The frequency of the oscillation is obtained with 
\begin{equation}
	f_{\mathrm{LF}} = \frac{\Delta A}{(L - K*\Delta A)}
	\label{equ:frequency_of_low_frequency_oscillation}
\end{equation}

Here we propose an improved LUT method. 
This LUT method starts with the same initial address for each output cycle.
The initial address is derived as follows by ensuring a zero-net value for each individual cycle. 
This requires the middle point of the lookup table results to be exactly at phase $\pi$.
For a sinusoidal LUT, suppose $K$ is odd, then 
\begin{equation}
    A\!\left(\frac{K - 1}{2}\right) = 0.
\end{equation}
If $K$ is even, then 
\begin{equation}
    A\!\left(\frac{K}{2}\right) + A\!\left(\frac{K}{2} + 1\right) = 0,
\end{equation}
i.e.,\
\begin{equation}
    A\!\left({\frac{K + 1}{2}}\right) = 0.
\end{equation}
If $\lfloor f_{\mathrm{clock}} / f_{\mathrm{o}} \rfloor$ is odd, the initial address should satisfy 
\begin{equation}
    A_0 + \frac{K - 1}{2} \Delta A = \frac{L}{2}.
\end{equation}
If even, it should instead satisfy
\begin{equation}
    A_0 + \left(\frac{K}{2} -1\right) * \Delta A + \frac{1}{2} \Delta A =  \frac{L}{2}.
\end{equation}
Combining both scenarios, we can obtain the universal expression of the initial address:
\begin{equation}
    A_0 = \frac{L}{2} - \frac{K-1}{2}\Delta A
\end{equation}
By applying the proposed LUT method, the reference signal can promise a net-zero value for any output cycle. 

The proposed LUT method can theoretically synthesize correct sinusoidal pulses over the full bandwidth.
However, the modulation method fails to cover the full frequency and amplitude range without causing output errors.   
Nearest-level modulation works by rounding the continuous reference signal to the nearest integer levels represented by multilevel output voltage steps, requiring only fundamental frequency switching:
\begin{equation} 
    s_{\mathrm{o, inte}} = \lfloor s_{\mathrm{in, }} \rceil
    \label{equ:nlm_principle}
\end{equation}
High-frequency reference signals allow only a limited number of modulation clock ticks within one output cycle. 
Since the advanced LUT method generate a reference signal that centers at the phase $\pi$, the reference signal is symmetric around $(\pi, 0)$. 
Theoretically, the digitized signal after NLM would still be symmetric and lead to a zero-sum for each individual output cycle. 
However, performing with modulation amplitude and limited length of representation for fixed-point variables in embedded controller can cause cases where the digitization still exposes a DC bias. 
Increasing the length of registers could reduce but will not eliminate the possibility of such implementation abnormality: 
\begin{align}
	\begin{split}
		\sum s_{\mathrm{ref, fp}} &= 0 \\
		\sum s_{\mathrm{mod, inte}} &\neq 0 \\
	\end{split}
\end{align}

\begin{figure}
    \centering
    \includegraphics{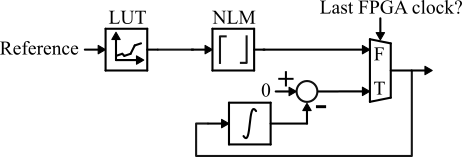}
    \caption{Diagram of the bias-eliminating nearest level modulation approach. 
    This modified NLM method integrates the rounding integers periodically, and feeds the result back to the output of NLM at the end of each output cycle, maintaining a same behavior as the conventional approach at low frequencies but avoiding distortion at high frequencies. }
    \label{fig:improved_nlm}
\end{figure}


We propose a modified NLM method that can adaptively eliminate the numerical error for high-frequency sinusoidal modulation.  
This modulation strategy ensures a zero accumulation value for each output cycle:
\begin{equation}
	s_{\mathrm{mod}, K} = 0 - \sum_{k=1}^{K-1} s_{\mathrm{mod}, k} 
	\label{equ:adaptive_hf_nlm}
\end{equation}
As per (\ref{equ:adaptive_hf_nlm}) and Fig.\ \ref{fig:improved_nlm}, the accumulation value is compared with zero at the end of the output cycle. 
The difference decides the last output command. 
Thus, the modulation can adaptively modify the output commands to achieve a net-zero value for each individual output cycle. 

\begin{figure*}[h]
    \centering
        \centering
        \includegraphics[width=\textwidth]{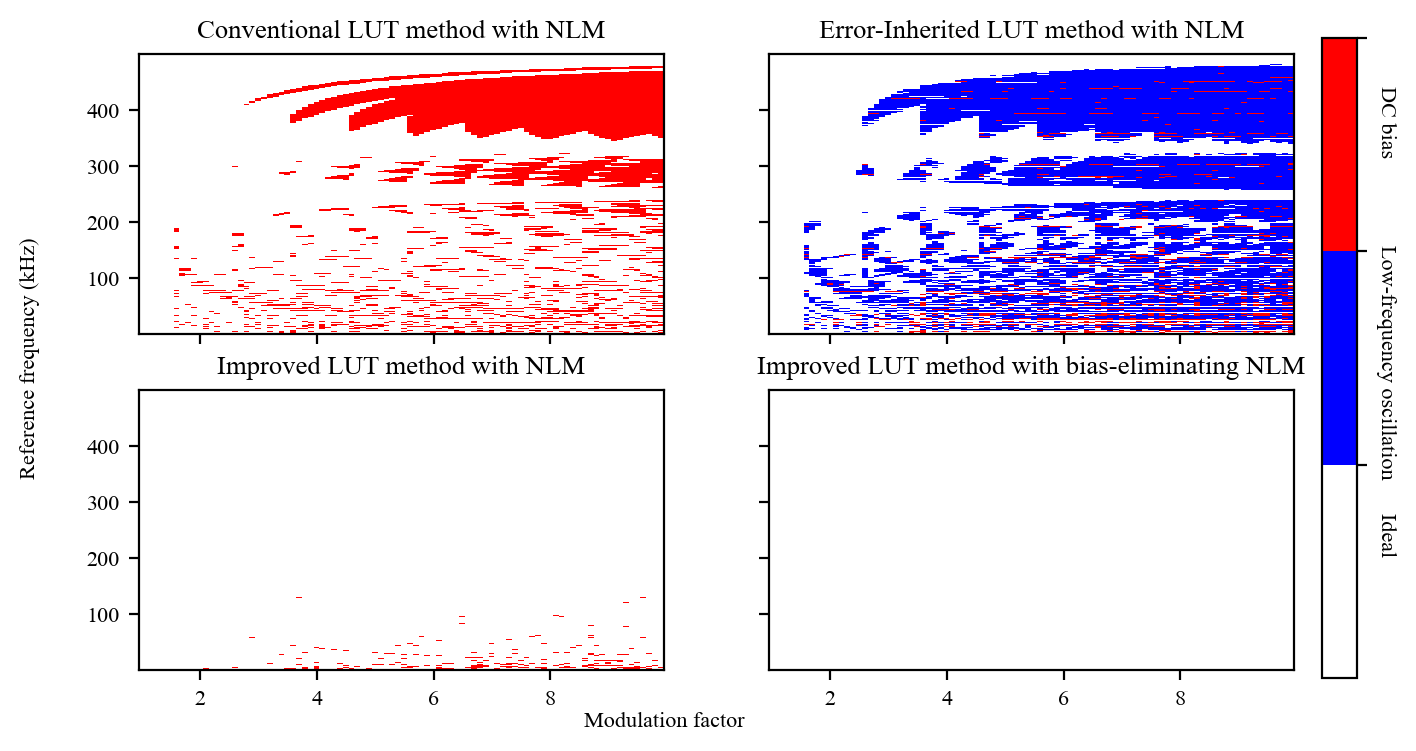}
	    \caption{
        Comparative analysis of four different LUT methods and modulation configurations: conventional LUT, error-inherited LUT, improved LUT with original NLM method, and improved LUT with adaptively-bias-eliminating NLM.
        The simulation encompasses a wide frequency range up to 5 MHz and an amplitude range to full modulation.
        Three output scenarios are observable: ideal performance, DC bias issue, and low-frequency oscillation issue.
        Conventional and error-inherited LUT methods exhibit notable DC bias or low-frequency distortion. The improved LUT method largely mitigates these issues, occasionally revealing minor imperfections. These rare imperfections are effectively addressed by the bias-eliminating NLM approach.
        }
    	\label{fig:modulation_verification}
\end{figure*}

\begin{figure}
    \centering
    \includegraphics{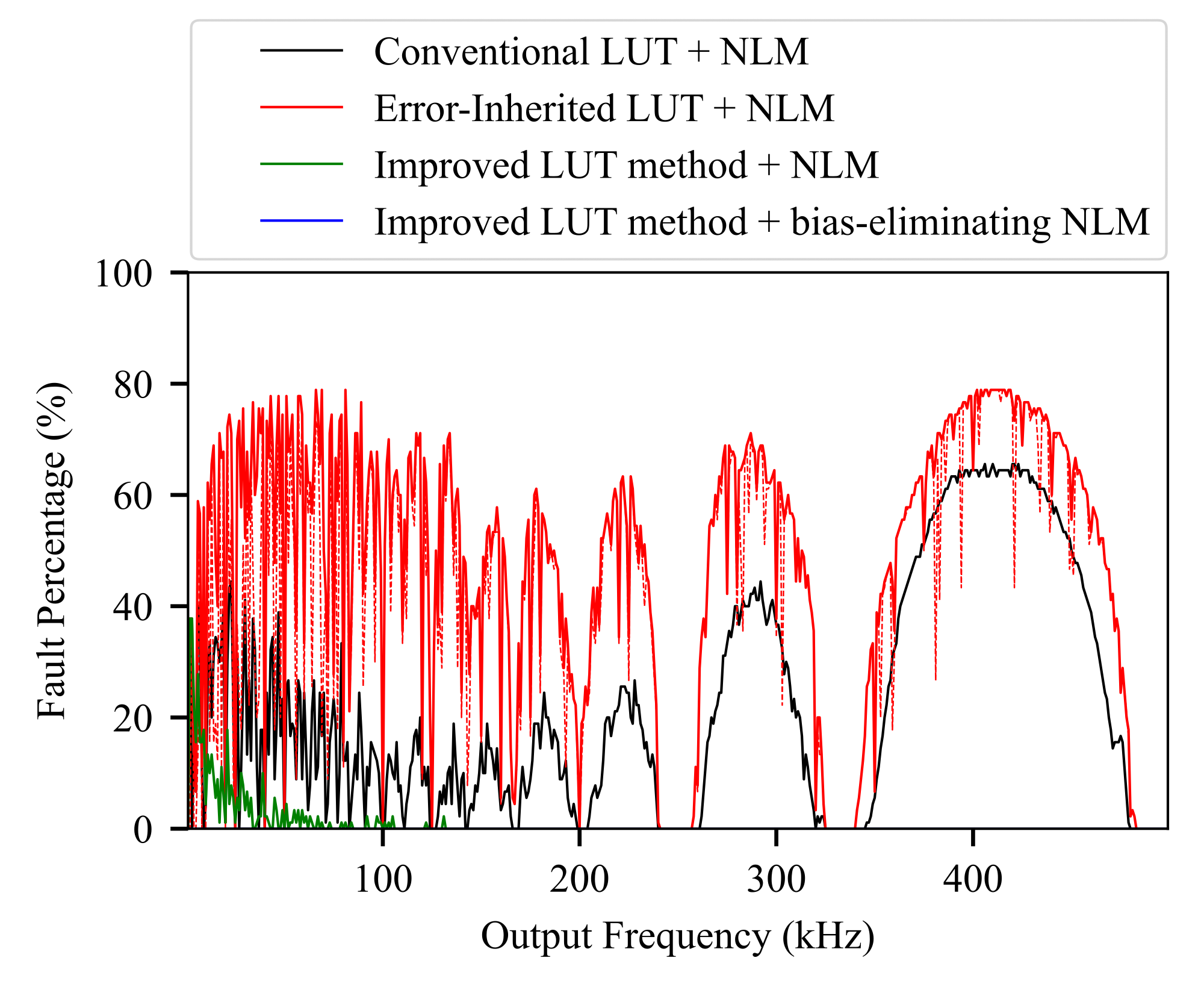}
    \caption{
        Quantitative perspective of modulation simulation results over output frequency, i.e., a quantitative presentation of results in Fig.\ \ref{fig:modulation_verification}, which intuitively compares the distribution of modulation performance over frequency and amplitude. 
        }
    \label{fig:quantized_simulation}
\end{figure}

In summary, the embedded control solves two crucial problems, respectively the synthetization of high-frequency sinusoidal signals with an improved LUT method, and their modulation with a modified NLM approach that can adaptively eliminate distortion.  

\section{Simulation Verification of Embedded Control Method}

The software aims for seamless operation across the entire bandwidth, spanning from DC to 5~MHz, while maintaining amplitude range integrity, without introducing a DC bias or significant distortion.

In our simulation, modulation factors ranged from 0 to 1 in increments of 0.01, and frequencies from 1~kHz to 5~MHz in 1~kHz steps. Four distinct software approaches were evaluated: zero-started LUT with NLM, error-inherited LUT with NLM, improved LUT with NLM, and improved LUT with adaptively-error-eliminating NLM.

The output current behavior can be categorized into three groups: the ideal case, DC bias, and low-frequency oscillation. DC bias and low-frequency oscillation both result in over-current situations, with the former leading to excessive current and the latter interfering with other frequency channels.

Figure \ref{fig:modulation_verification} compares these software implementations. The conventional LUT method displays prominent DC bias, particularly at higher frequencies. The error-inherited LUT method is susceptible to low-frequency oscillation, especially at elevated frequencies. The improved LUT method significantly mitigates DC bias and low-frequency oscillation, though due to embedded controller limitations, a DC bias risk remains. Conversely, the adaptively-error-eliminating NLM method ensures optimal output results across the entire frequency and amplitude range, satisfying the prototype's performance criteria.

Figure \ref{fig:quantized_simulation} provides a statistical overview of modulation outcomes. Existing LUT methods exhibit distortion across a broad frequency range. The proposed LUT method effectively addresses most issues, except for minor defects at lower frequencies. Incorporating the bias-eliminating NLM approach successfully resolves modulation challenges.

\section{Experimental Results}
\subsection{Setup}
In this section, we present the high-power and wide-bandwidth capability with frequency sweeping, i.e., a chirp signal, and showcase the prototype with variously featured output profiles. 

\begin{table}[t]
    \caption{Key components in prototype design}
    \centering
    \begin{tabular}{c|c}
       \hline
       Item  & Product Information\\
       \hline
       Transistor  & GS61008T (100V, 7 $\rm{m\Omega}$, GaN Systems)\\
       Transistor driver IC & ACPL-P346-000E (2.5 A, BroadCom)\\
       Power supply & HP 6030A (200V/17A, 1200 W, Agilent)\\
       Embedded controller & sbRIO-9627 (Zynq-7020 FPGA, National Instruments)\\
       \hline
    \end{tabular}
    \label{tab:experiment_platform_summary}
\end{table}

\begin{table}[t]
    \caption{Measurement tools}
    \centering
    \begin{tabular}{c|c}
       \hline
       Item  & Product Information\\
       \hline
       Oscilloscope  & MDO3054 (500 MHz, 2.5 GSPS, Tektronix)\\
       Voltage probe & THDP0100 (100 MHz, 2.3 kV, Tektronix)\\
       Current probe  & HF60R (30 MHz, CWTMini, PEM)\\
       Spectrogram program  & matplotlib.pyplot.specgram (Python Package)\\
       \hline
    \end{tabular}
    \label{tab:measurement_summary}
\end{table}

\begin{figure}[t]
    \centering
    \includegraphics[width=0.45\textwidth]{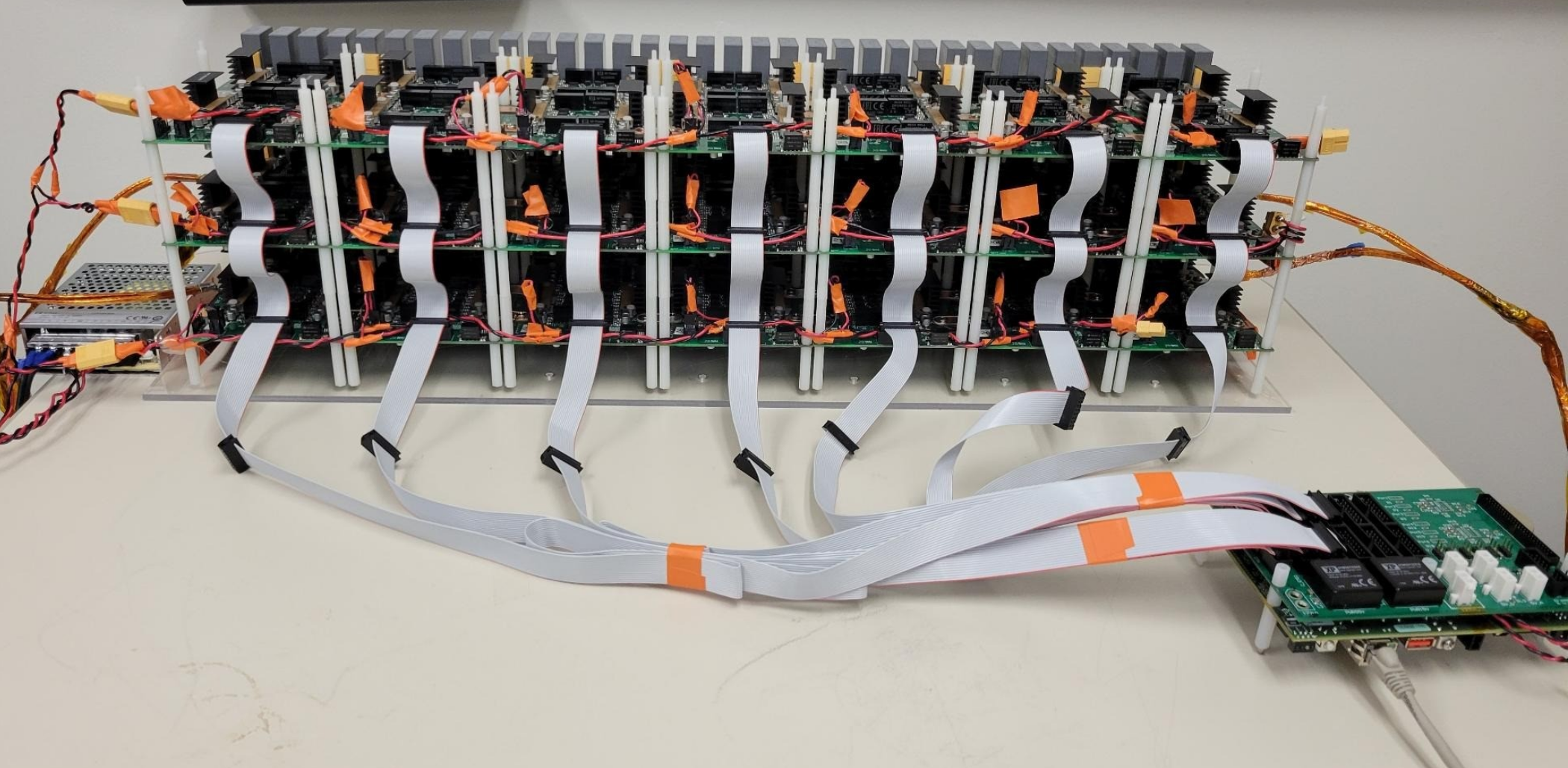}
	\caption{
    High-power wide-bandwidth prototype hardware of the circuit.
    }
    \label{fig:prototype_photo}
\end{figure}

We designed and constructed the prototype as illustrated in Figure \ref{fig:prototype_photo} with key parameters as per Table \ref{tab:experiment_platform_summary} and tools from Table \ref{tab:measurement_summary} to demonstrate the key improvements over the state of the art, specifically achieving high quality of the current output (indicated by the distortion), e.g., needed to drive magnetic fields, and wide-bandwidth output through switching power electronics with its high power and efficiency capabilities.




%

\subsection{Sweeping-Frequency Output: Characterizing Power and Bandwidth Capability as well as Fidelity}

\begin{figure}
    \centering
    \includegraphics[width=0.45\textwidth]{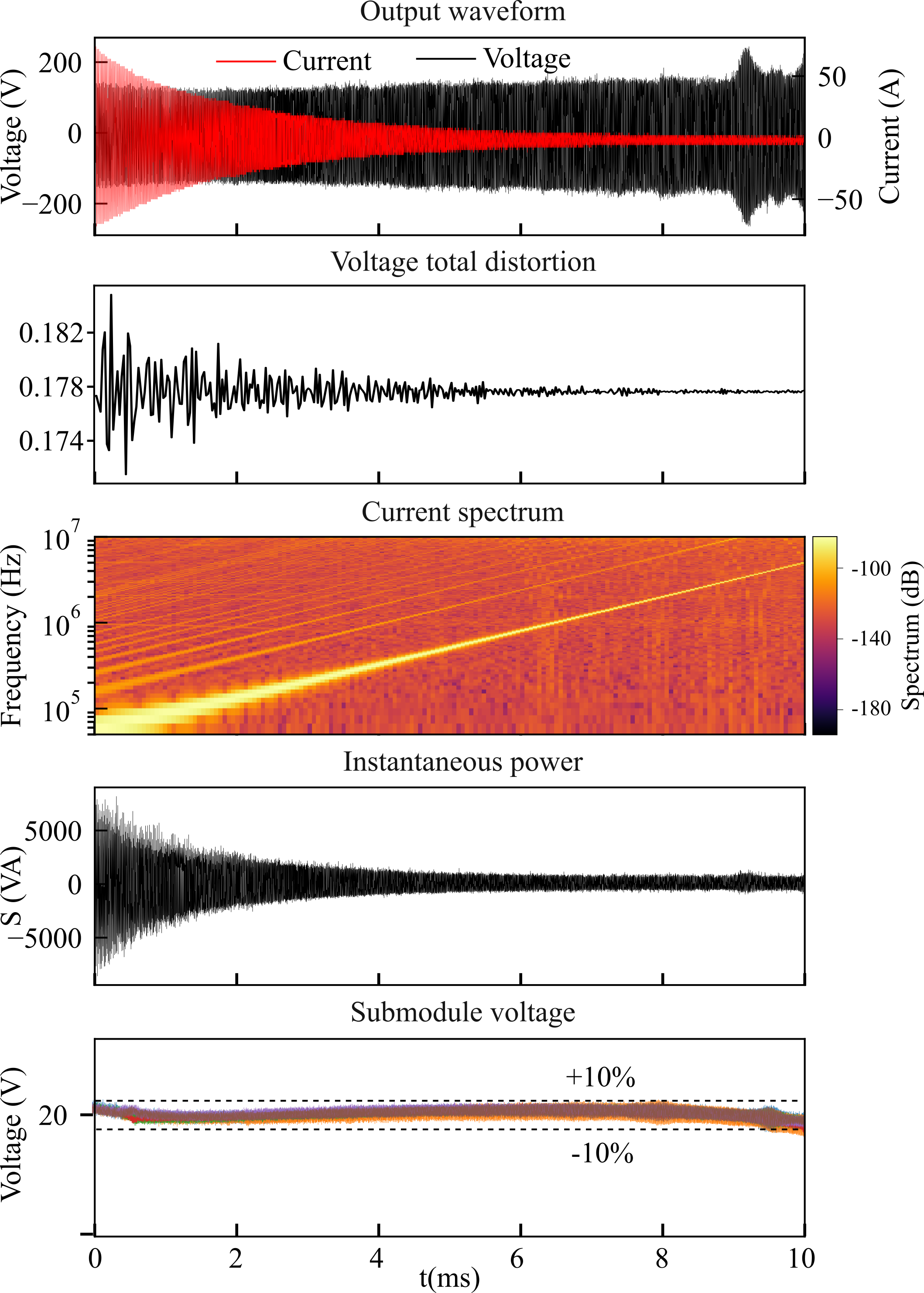}
	\caption{
    Frequency sweeping demonstrated with a chirp signal that exponentially increases frequency from DC to 5~MHz within 10~ms.
    }
    \label{fig:chirp}
\end{figure}

Figure \ref{fig:chirp} demonstrates the spectrogram of a frequency sweep from DC to 5~MHz in 1~ms, while maintaining a widely constant voltage amplitude at full modulation factor. The peak voltage level of 140~V corresponds to the sum of the module voltages ($7\times 20$~V). 
Due to the inductive load, the current is decreasing from initially some 50~A with rising frequency. 

The total voltage distortion remains consistently below 18.4\%, with an average distortion of 17.74\% across the frequency spectrum.
\marginpar{\tiny{Either provide quantitative statements about the current (distortion) or remove it.}}
Due to switching, most distortion is in regular harmonics as the parallel lines in the current spectrogram demonstrate.
The balancing solution keeps the voltages of the submodules within a range of $\pm 2$ V (equivalent to $\pm 10\%$).


\subsection{Frequency Mixing}

\begin{table}[h]
    \caption{Summary of Channel Mixture Presentation}
    \centering
    \begin{tabular}{c|c|c}
       \hline
	    Frequency  & Normalized voltage amplitude & Interval\\
       \hline
	    100~kHz  & 2 & $0   \sim 200~\rm{\mu s}$ \\
	    200~kHz  & 4 & $100 \sim 500~\rm{\mu s}$ \\
	    400~kHz  & 3 & $400 \sim 800~\rm{\mu s}$ \\
	    1~MHz    & 4 & $600 \sim 700~\rm{\mu s}$ \\
	    3~MHz    & 4 & $700 \sim 800~\rm{\mu s}$ \\
	    5~MHz    & 6 & $800 \sim 1000~\rm{\mu s}$ \\
       \hline
    \end{tabular}
    \label{tab:summary_channel_mixture}
\end{table}

\begin{figure}
    \centering
	\includegraphics[width=0.45\textwidth]{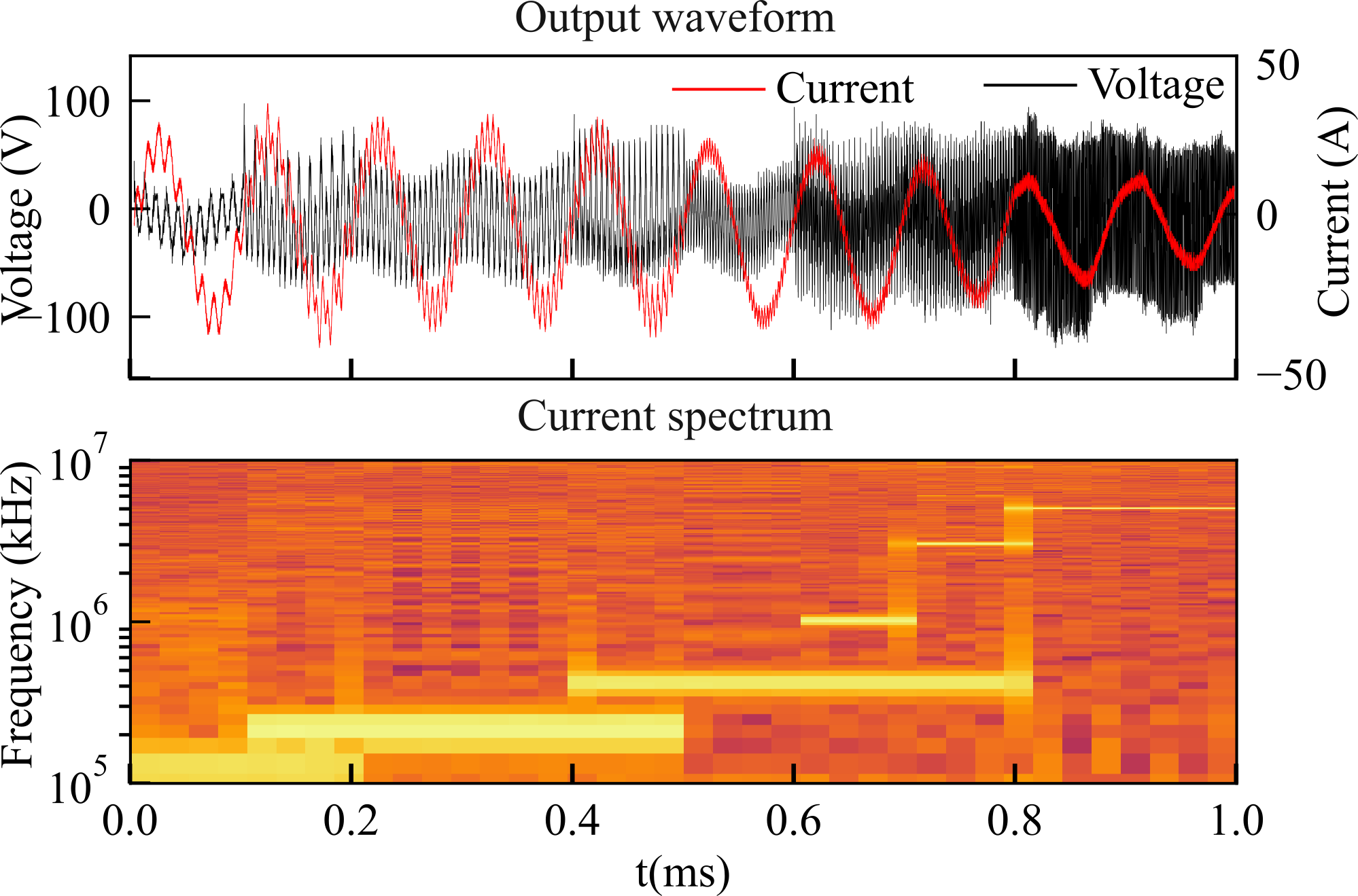}
	\caption{Neuronal stimulation oriented channels-mixture trial. 7 channels are selected to activate potential magnetic nanoparticles. Each channel is activated with specified amplitude and interval.}
    \label{fig:channels_mix}
\end{figure}

\begin{figure}
    \centering
    \includegraphics[width=0.45\textwidth]{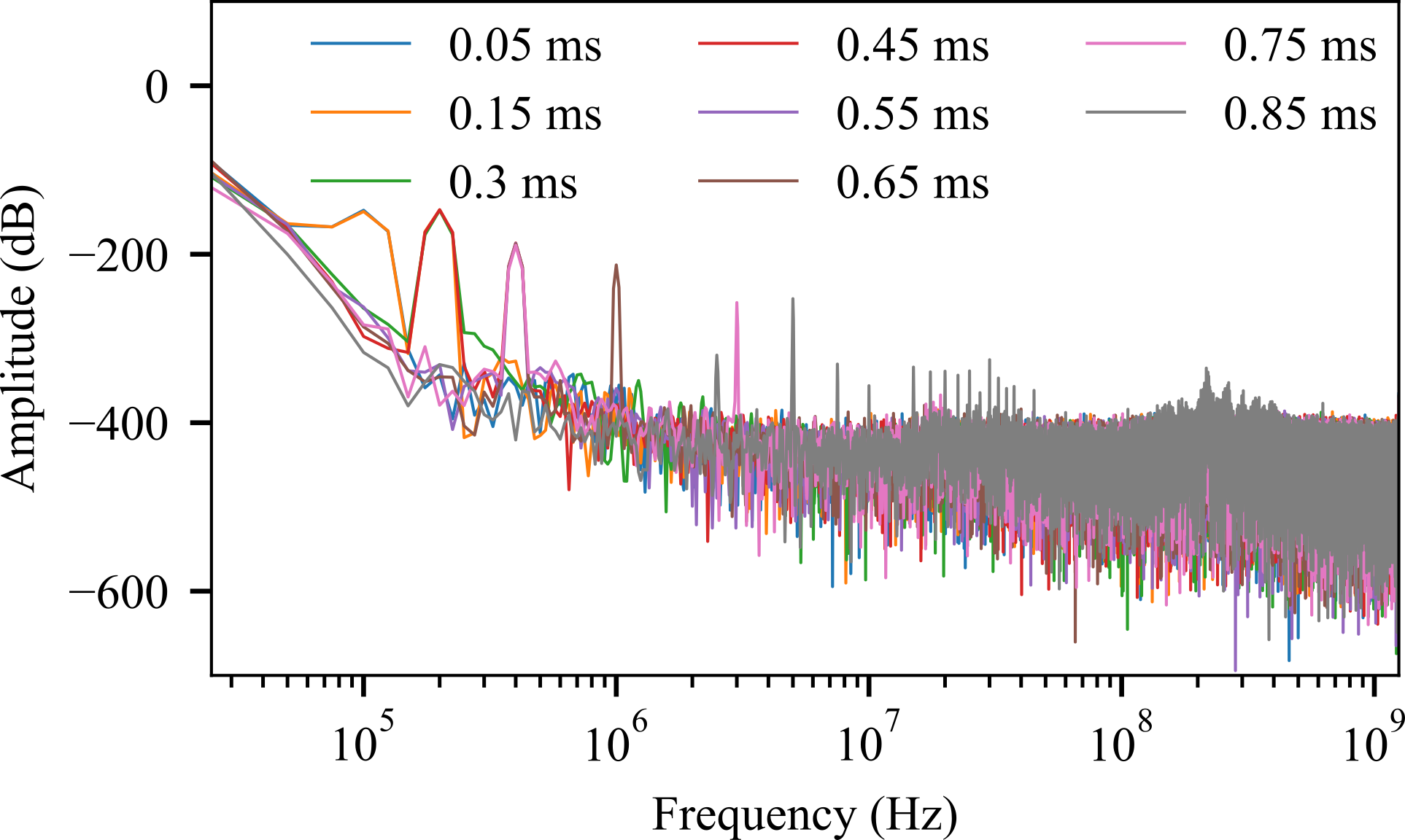}
    \caption{Fourier transformation analysis of the experiment in Fig. \ref{fig:channels_mix} at moments that cover all diverse combinations of frequency channels. }
    \label{fig:channels_mix_fft}
\end{figure}

Various applications, such as multichannel wireless power transfer, inductive power encryption, and magnetogenetics for various frequency-selective nanoparticles require the independent generation and control of several frequencies, including independent amplitude variation and frequency modulation.
%
Figure \ref{fig:channels_mix} demonstrates such frequency-division signals in voltage, current, and spectrogram with the components listed in Table \ref{tab:summary_channel_mixture}.
%
The fundamental mode at 10~kHz is maintained, while higher-frequency channels are mixed in. 


The spectra of each interval of Fig.\ \ref{fig:channels_mix} lead to sharp lines in Fig.\ \ref{fig:channels_mix_fft}, only broadened by the window lengths.

\subsection{Arbitrary Multifrequency Output}

\begin{figure}
    \centering
	\includegraphics[width=0.45\textwidth]{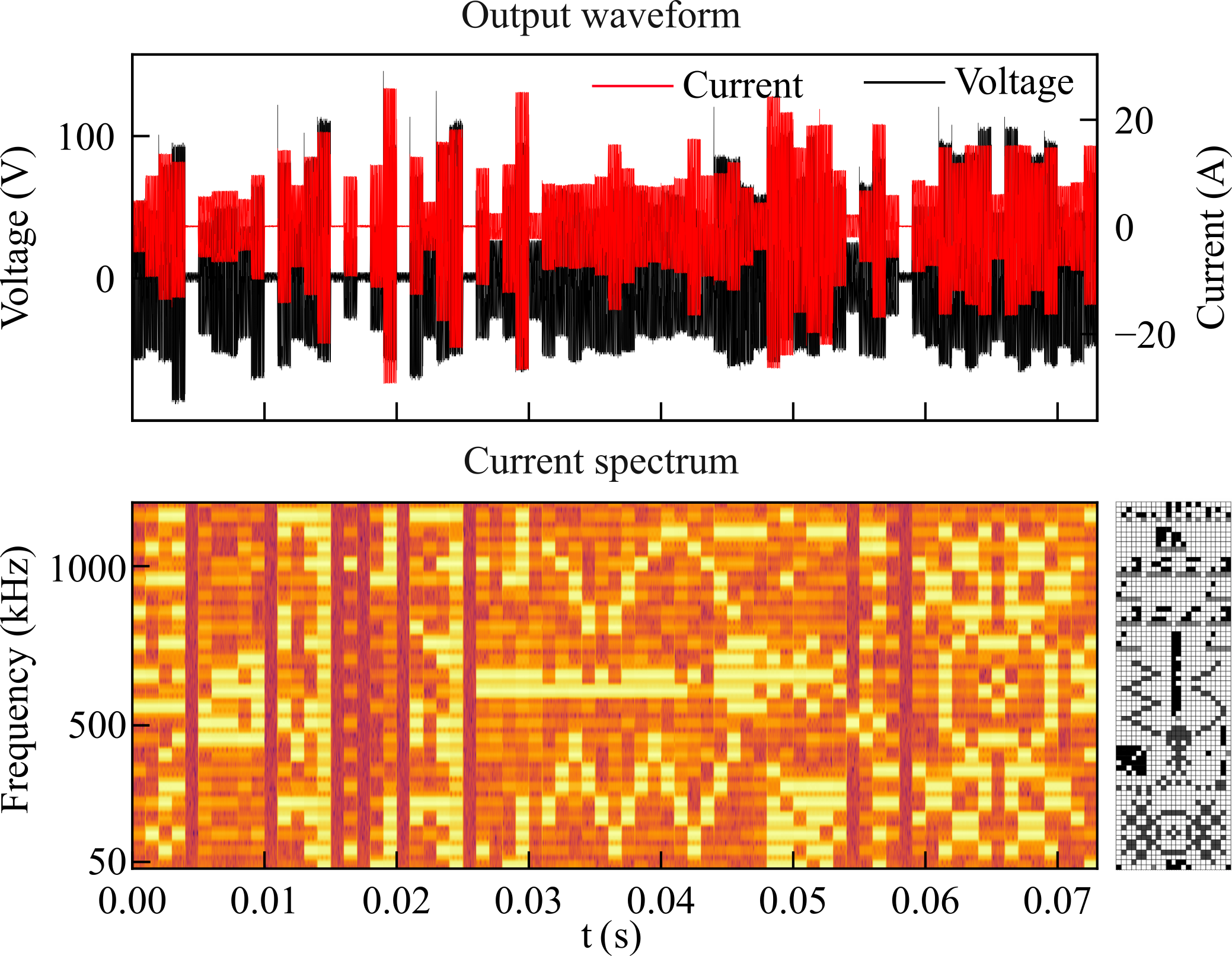}
	\caption{Arecibo reply message is compressed in the current spectrogram. 23 frequency channels, respectively 50~kHz, 100~kHz, linearly increasing to 1150~kHz with 50~kHz step size, are used to reproduce the message within 73 ms, representing 23 x 73 pixels in the original message.
    } 
    \label{fig:arecibo}
\end{figure}


The circuit with the control can modulate the signal fast enough to serve as a signal amplifier. Figure \ref{fig:arecibo} displays the measurement of the Arecibo message sent through our magnetic antenna \cite{arecibo1974}. 




Furthermore, this technique holds promise for encrypted communication. By employing customized sequences of frequency channels instead of linearly natural ones, the received information or image would resemble a confounding puzzle, akin to having undergone a paper shredder. Only when the correct frequencies and sequence are supplied does the received information coalesce into a coherent form.

\section{Conclusion}

\begin{figure}
    \centering
	\includegraphics[width=0.45\textwidth]{img/prototypes_capability_bar.pdf}
	\caption{Prototype capability demonstration compared with other power converters. } 
    \label{fig:prototypes_capability}
\end{figure}

This paper demonstrated the generation of concurrent high power and high bandwidth with high quality with a combination of several measures.  The power--bandwidth product exceeded 25~$\mathrm{GV\cdot A \cdot Hz}$, which exceeds the state of the art, whose figures lie $10\sim 100~\mathrm{MV\cdot A \cdot Hz}$, by more than 2 orders of magnitude, and a total distortion (or THD+N) of ~17\%. In contrast to monolithic circuits, the cascaded design allows further scaling for even higher levels. 

As no single ingredient alone was able to achieve the target values, we combined a cascaded bridge circuit, fast-switching GaN transistors, and proper embedded control. Modules allowing dynamic series and parallel inter-module connectivity enabled stable and fast sensor-less module balancing.

We demonstrated the prototype's abilities to generate rapidly changing wide-bandwidth signals in the kilowatt range.

The solution overcomes the previous limitation of concurrent power, bandwidth, and quality. Whereas conventional monolithic circuits, both resonant and hard-sitching, have a fixed product of these values, the cascaded topology allows scaling all three of them by adding additional modules that share power, improve the quantization granuarity, and spread the necessary switching speed out.

Various applications from multi-channel inductive power transfer, wireless energy encryption, combined power and signal transmission to medical implants, and magnetogenetics may immediately benefit from the development.


\section*{Acknowledgment}
This research was developed with funding from the Defense Advanced Research Projects Agency (DARPA) of the United States of America, Contract No. N66001-19-C-4020. The presented views, opinions and/or findings are those of the authors and should not be interpreted as representing the official views or policies of the Department of Defense or the U.S.A. Government.

\bibliographystyle{Bibliography/IEEEtranTIE}
\bibliography{Bibliography/IEEEabrv,Bibliography/main}\ 

\end{document}